# High-dimensional entanglement between distant atomic-ensemble memories

Dong-Sheng Ding[1,2,†,§], Wei Zhang[1,2,†], Shuai Shi[1,2,†], Zhi-Yuan Zhou[1,2], Yan Li[1,2], Bao-Sen Shi[1,2,*] and Guang-Can Guo[1,2]

**Entangled quantum states in high-dimensional space show many advantages compared with entangled states in two-dimensional space. The former enable quantum communication with higher channel capacity, enable more efficient quantum-information processing, and are more feasible for closing the detection loophole in Bell test experiments. Establishing high-dimensional entangled memories is essential for long-distance communication, but its experimental realization is lacking. We experimentally established high-dimensional entanglement in orbital angular momentum space between two atomic ensembles separated by 1 m. We reconstructed the density matrix for a three-dimensional entanglement and obtained an entanglement fidelity of 83.9±2.9%. More importantly, we confirmed the successful preparation of a state entangled in more than three-dimensional space (up to seven-dimensional) using entanglement witnesses. Achieving high-dimensional entanglement represents a significant step towards a high-capacity quantum network.**

**Keywords**: high-dimensional entanglement, orbital angular momentum, quantum memory

[1]Key Laboratory of Quantum Information, University of Science and Technology of China, Hefei, Anhui 230026, China

[2]Synergetic Innovation Center of Quantum Information and Quantum Physics, University of Science and Technology of China, Hefei, Anhui 230026, China

[†]These authors contributed to this article equally

Correspondence: [§]Ding D-Sdds@ustc.edu.cn; [*]Shi B.S.drshi@ustc.edu.cn

# INTRODUCTION

Quantum entanglement distributed in different nodes is essential for realizing long-distance quantum communications [1]. By introducing a quantum repeater protocol, the problem of the exponential scaling of the error rates with channel length in quantum communication can be overcome by using entanglement storage and swapping operations [2]. Usually, the stored photons are encoded in a two-dimensional space, such as polarization, which-path, and time-bin, which results in information being carried by a photon as a qubit. Recently, another photonic degree of freedom, orbital angular momentum (OAM) [3–5], has attracted much interest because the OAM states of a photon could belong to a high-dimensional space, which would enable encoding with inherent infinite degrees of freedom, thereby enhancing the channel capacity and significantly improving the efficiency of a network [6, 7].

Constructing quantum networks based on OAM involves the preparation of high-dimensional entangled photons and the realization of high-dimensional entangled memories. The repeaters for a high-dimensional quantum communication network are based on many high-dimensional entangled memory modes, and long-distance communication can be realized via a swapping operation between adjacent entangled memories. Thus, establishing high-dimensional entangled memories is a critical step. Progress has been made in preparing high-dimensional OAM entangled states, including the confirmation of a three-dimensional entanglement between a delayed atomic spin wave and a photon [8] prepared in spontaneous Raman scattering (SRS), an 11-dimensional entanglement [9], a high-dimensional image entanglement [10], and a 100×100 OAM entanglement [11] between two photons generated by spontaneous parametric down-conversion. In recent years, many groups have explored the realization of high-dimensional entangled memories with some success [12–23], but all work to date has only displayed approaches to this goal. The experimental works on establishing entangled memories using different physical systems have focused on the entanglement in a two-dimensional space [24–29]. Until now, the storage of a high-dimensional OAM state on-demand in any physical system, which is essential for the realization of quantum repeaters for long-distance quantum communications, and the establishment of a high-dimensional OAM entanglement between different quantum memories have not been reported. Trying to solve this problem is the main motivation for this work. A previous work [24] reported the creation of a two-dimensional OAM entanglement between two atomic ensembles, but the realization of a high-dimensional entanglement between different quantum memories is non-trivial and is not a straightforward extension of establishing a two-dimensional entanglement between two atomic ensembles. There are many challenges, such as in proving the high-dimensional entanglement and determining the dimensionality of entanglement. Usually, we can characterize storage based on entanglement fidelity by comparing the density matrixes before and after storage, which are reconstructed via quantum tomography. For a high-dimensional entangled state, the number of measurements required to reconstruct a density matrix scales quadratically with dimensionality. Therefore, reconstructing the density matrices for a high-dimensional (>3) entangled state is impractical because the amount of data that must be measured increases significantly. Moreover, balancing the distinct efficiencies when creating and storing different OAM modes in a high dimension is a challenge, which becomes more serious with increasing dimensions.

In this study, we report the experimental establishment of high-dimensional OAM entanglement between two quantum memories, making a primary step towards the building of a high-dimensional quantum network. In our experiment, we entangled the high-dimensional OAM states of a photon and a collective spin-excited state in a cold atomic ensemble via SRS [8] and then sent the photon to be stored in another cold atomic ensemble using the Raman protocol [30]. In this way, we established high-dimensional OAM entanglement between two atomic ensembles. We confirmed the entanglement by mapping the spin-excited states in the two ensembles to two photons and checked their entanglement. We reconstructed the density matrices of the three-dimensional OAM entangled photons. The entanglement fidelity was calculated to be 83.9±2.9%. We then used an entanglement witness to characterize the nature of the higher dimensional entanglement and concluded that there is at least a four-dimensional entanglement within the two memories. Using the dimensionality witness, we confirmed a seven-dimensional entanglement between these memories.

**MATERIALS AND METHODS**

The experimental media were optically thick atomic ensembles of Rubidium 85 ($^{85}$Rb) that were trapped in two two-dimensional magneto-optical traps (MOT) [30] and were separated by 1 m but worked independently. A schematic of the energy levels involved and the experimental setup are shown in Fig. 1(a) and (b). By inputting Gaussian pulse modes with a pulse width of 30 ns, we generated in MOT 1 via SRS an OAM entanglement between the signal-1 photon and the collective spin-excited state of the atomic ensemble. In this process, the laser pulse of pump 1 was blue-detuned by 70 MHz to the atomic transition $|3\rangle$ ($5S_{1/2}(F=3)$)→$|2\rangle$ ($5P_{1/2}(F'=3)$) and signal-1 was blue-detuned by 70 MHz to the atomic transition $|2\rangle$ ($5P_{1/2}(F'=3)$)→$|1\rangle$ ($5S_{1/2}(F=2)$). Because this is a SRS process that conserves momentum, the initial state of the system has zero linear and angular momentum; thus, the resulting joint state of the signal-1 photon and the atomic spin-excited state has zero total angular momentum, which induces OAM correlations between them. The established entanglement is written as $|\psi\rangle = \sum_{m=-\infty}^{m=\infty} c_m |m\rangle_{s1} \otimes |-m\rangle_{a1}$, where $|m\rangle$ denotes the OAM state of the $m$ quantum eigenmode; the subscripts $s1$ and $a1$ refer to the signal-1 photon and the atomic ensemble in MOT 1, respectively; and $|c_m|^2$ is the excitation probability for different OAM modes. Next, the signal-1 photons were sent to be stored as a collective spin-excited state of the atomic ensemble trapped in MOT 2 using the Raman scheme [31]. Hence, the two atomic ensembles in different MOTs are in a high-dimensional entanglement of $|\psi'\rangle = \sum_{m=-\infty}^{m=\infty} o_m |-m\rangle_{a1} \otimes |-m\rangle_{a2}$, where $|o_m|^2$ is the amplitude probability for the different modes $m$ and subscript $a2$ refers to the atomic collective spin-excited states in MOT 2. The power of the coupling laser is 40 mW, with a beam waist of 2 mm, corresponding to a Rabi frequency of 10.6 $\Gamma$ ($\Gamma$ is the decay rate of level $5P_{1/2}(F'=3)$). The storage time T2 in MOT 2 should not be longer than the storage time of the spin wave T1 in MOT 1 to demonstrate the storage of the entanglement. In our experiment, the condition T2=T1=100 ns was taken, i.e., the coupling laser and the pump-2 laser were opened at the same time. Furthermore, we applied the method detailed in Ref. [31] to match the bandwidth between the signal-1 photons and the memory for high storage efficiency, thus enabling a storage efficiency of 26.8% to be achieved experimentally. The memory efficiency was calculated as the ratio of the coincidence counts before and after storage. After a delayed time of 100 ns, we used the pump-2 laser with a square pulse width of 250 ns, resonant with the atomic transition $|1\rangle$ ($5S_{1/2}(F=2)$)→$|4\rangle$ ($5P_{3/2}(F'=3)$), to read the spin wave to generate the signal-2 photon, which corresponds to the transition $|4\rangle$ ($5P_{3/2}(F'=3)$) →$|3\rangle$ ($5S_{1/2}(F=3)$). Simultaneously, we switched the coupling laser on to read the signal-1 photon emitted out of the atomic ensemble in MOT 1. The delay time was measured by comparing the peaks of the two-photon coincidence with and without delayed pumps 1 and 2. Pumps 1 and 2 had a power of 0.5 mW and 4 mW, respectively. A check of the entanglement between the signal-1 and signal-2 photons verified the high-dimensional entanglement between the atomic ensembles.

## RESULTS AND DISCUSSION

Before verifying the high-dimensional entanglement, the correlations in the OAM space between the two ensembles before and after storage were first measured. The results are shown in Fig. 2, with panels (a)/(b) showing the OAM correlations between photons of signal 1 and signal 2 without/with storing signal 1. This also characterizes the correlations between the signal-1 photon and the atomic collective spin-excited state in MOT 1. The differences in OAM correlations shown in Fig. 2(a) arise from different OAM distributions in the nonlinear SRS process [Fig. 2(c)]. Because the efficiencies were different when storing the various OAM modes $|m\rangle$, the resulting correlation matrix after storage [Fig. 2(b)] was slightly different from that before storage [Fig. 2(a)]; this is similar to the result that was obtained when using a weak coherent light [6]. The different storage efficiencies measured for different OAM modes $m$ are shown in Fig. 2(d).

Next, we verified the entanglement with $d=3$. We projected the signal-1 and signal-2 photons onto SLM 1 and SLM 2, respectively, with nine different phase states $|\psi_{1-9}\rangle$ corresponding to states $|L\rangle$, $|G\rangle$, $|R\rangle$, $(|G\rangle+|L\rangle)/2^{1/2}$, $(|G\rangle+|R\rangle)/2^{1/2}$, $(|G\rangle+i|L\rangle)/2^{1/2}$, $(|G\rangle-i|R\rangle)/2^{1/2}$, $(|L\rangle+|R\rangle)/2^{1/2}$, and $(|L\rangle+i|R\rangle)/2^{1/2}$ [6, 32], where $|L\rangle$, $|G\rangle$, and $|R\rangle$ are states corresponding to a well-defined OAM of $-\hbar$, 0, and $+\hbar$, respectively. With this projection, the mirrors and lens should be considered due to the photon's transformation during the imaging process. We reconstructed the density matrix before storage [Fig. 3(a) and (b)] by converting the spin-excited state in MOT 1 into a signal-2 photon. The signal-1 photon was then stored for a while in MOT 2 following a Raman protocol. This established the entanglement between the two atomic ensembles, for which the reconstructed density matrix is given in Fig. 3(c) and (d). The difference between the states before and after storage was due to the different storage efficiencies for the $|R\rangle$ ($|L\rangle$) mode and $|G\rangle$ mode [Fig. 2(d)]. To check the entangled state before and after storage, we used pump 2 and the coupling lasers to read the atom–atom entangled state into the retrieved signal 1–signal 2 entangled state and then checked the entanglement between these photons. The reconstructed density matrices are shown in Fig. 3(a)–(d). Using the formula $F_1 = Tr(\sqrt{\sqrt{\rho_x}\rho_{ideal}\sqrt{\rho_x}})^2$, where $x$ represents the input and output and $\rho_{ideal}$ is the density matrix of the ideal three-dimensional OAM entangled state of $|\Psi_{ideal}\rangle = (|R\rangle_{a1}|R\rangle_{a2} + |G\rangle_{a1}|G\rangle_{a2} + |L\rangle_{a1}|L\rangle_{a2})/3^{1/2}$, we calculated the fidelity of the reconstructed density matrix before and after storage, which was 76.7%±2.8% and 71.7%±2.8%, respectively. All error bars in this experiment were estimated using Poisson statistics and performing Monte Carlo simulations using Mathematica software. Both exceeded the threshold of 2/3 [8, 33] for a maximally entangled state of Schmidt rank 3, which confirms that the density matrix cannot be decomposed into an ensemble of pure states of Schmidt rank 1 or 2, i.e., the Schmidt number of the density matrix must be equal to or greater than 3 both before and after storage. We calculated the fidelity of entanglement with $F_2 = Tr(\sqrt{\sqrt{\rho_{output}}\rho_{input}\sqrt{\rho_{output}}})^2$, which yielded 83.9%±2.9%.

Finally, we focused on the main part of this study, i.e., the establishment of a high-dimensional entanglement between two memories. In principle, the density matrices of the higher-dimensional entanglement can be reconstructed using the above method, but in practice, there are some experimental challenges in its realization. For example, for a $d$-dimensional entangled state, the amount of data needed is of the order $d^4$, which makes the reconstruction of the density matrix impractical. Basically, there are three methods for checking whether a system is in high-dimensional entanglement: 1) using unbiased basis states that span the whole subspace [34, 35]; 2) checking inequalities in higher dimensions directly [36, 37]; or 3) finding a violation that is stronger than allowed within a two-dimensional state space, thereby hinting at entanglement in (untested) higher dimensions. Here, we used method 3 to characterize the entanglement. We used the entanglement witness [38, 39] to prove whether there was a high-dimensional entanglement and the dimensionality witness [40–42] to characterize the dimensionality of the entanglement.

These witnesses determine the level of entanglement using a minimum number of measurements. To calculate the witnesses, we only need to measure all of the states entangled in a two-dimensional OAM subspace, i.e., the correlations in three mutually unbiased bases, including diagonal/anti-diagonal, left/right, and horizontal/vertical bases, need to be measured, with the number of data points needed being reduced to $3d(d-1)$ [43]. For example, to reconstruct the density matrix, the amount of data that must be measured for 3D is 81, and for 4D, it is 256. If we use the witness, then the amount of data needed is reduced significantly to 18 and 36, respectively. This greatly shortens the experimental time. The entanglement and dimensionality witnesses can be calculated from the sum of the visibilities $M = V_x + V_y$ and $N = V_x + V_y + V_z$, respectively, in each 2×2 subspace, where the visibilities are defined as $V_i = |\langle \sigma_i \otimes \sigma_i \rangle|$, $i = x, y, z$. Here, $\sigma_x, \sigma_y, \sigma_z$ represent the measurements in the diagonal/anti-diagonal, left/right, and horizontal/vertical bases, respectively. The superposition is calculated by adding equal amounts of the two modes, and the phase is calculated just from the argument of the resultant complex [44]. Figure 4(a) shows an example of mutually unbiased bases formed from the OAM modes $m=5$ and $m=-1$. For a separable state within a $d$-dimensional subspace, a product state of a $(d-1)$-dimensional maximum entangled state and a single state $\psi_{system} = \psi_{d-1} \otimes \psi_1$ maximizes the sum of the visibilities. Because the allowed maximum visibility of entanglement in a two-dimensional subspace is 2 ($M = V_x + V_y = 2, V_x = V_y = 1$), the allowed maximum visibilities can be calculated as $(d-1)(d-2)$ for a $(d-1)$-dimensional entanglement. The maximum visibilities for the remaining separable state are $(d-1)$ [39]. Hence, the maximum bound for high-dimensional entanglement is given as

$$M_d = (d-1)^2. \qquad (1)$$

If there is a $d$-dimensional entanglement, the maximum bound of $M_d$ should be violated. For a state comprising $m=2, 1, 0, -1$, the maximum bound is $M_4=9$. The measured $M'$ is 9.30±0.06 and 9.19±0.06 before and after storage, respectively. These values clearly suggest that there is at least a four-dimensional entanglement between these distant memories.

By assuming the correlations $|\psi\rangle = \sum_{m=-\infty}^{m=\infty} c_m |m\rangle_{s1} \otimes |-m\rangle_{a1}$, as in Ref. [24], we also sum the visibilities $N$ for each of the bases to calculate a witness value $W$ to determine the dimensionality of high-dimensional entanglement. All experimentally measured visibilities $N$ are shown in Fig. 4(b) and (c), corresponding to quantities before and after storage, respectively. The dimensional witness value [24] is given by

$$W_d = 3\frac{D(D-1)}{2} - D(D-d), \qquad (2)$$

where $D$ is the number of OAM modes in the measurement. If $W>W_d$ holds, the memories are entangled in at least $d+1$-dimensions. In our experiment, the measured number of modes was 11 ($m=-5\rightarrow5$); the obtained $W$ of 123.9±0.8 for the input state and 112.8±0.8 for the output state violated the bound of 110 for an input of $d=6$ and 99 for an output of $d=5$, both by 17 standard deviations, implying that there strongly exists a six-dimensional entanglement between the memories. The obtained $W$ also violated the bound of 121 for an input of $d=7$ and 110 for an output of $d=6$, both by 3 standard deviations, demonstrating that there exists a seven-dimensional entanglement and indeed a high-dimensional entanglement in our memories.

In demonstrating three-dimensional entanglement, the fidelity was affected by the distinct storing efficiencies [Fig. 2(d)] for different OAM modes [6], narrowing the spiral bandwidth of the OAM modes. We can improve the entanglement fidelity by purifying the entangled state [10, 45]. However, achieving a balance in storing different OAM modes is a big challenge [46]. In all experiments, entanglement was verified by checking the entangled photon readout from the atomic ensembles, which were found to be a-posteriori entangled. Both signal photons were completely covered by the pump and coupling laser beams in our experiment; hence, it is reasonable to assume that the readout efficiencies for different OAM modes are the same. Hence, the photonic entangled state can be regarded as a post-selected entangled state of the atomic ensembles.

The main reasons affecting the entanglement dimension are as follows: 1) the distinct efficiencies in storing the different OAM modes; 2) the noise associated with storage, which is mainly from the scattering from the coupling laser, resulting in a low signal-to-noise ratio (SNR); 3) the stability of the whole system over long experimental periods (the total measured time for the storage and retrieval processes was approximately 100 h). To increase the OAM entanglement dimension, we believe four problems need to be solved: a) generating a maximal high-dimensional entanglement between the signal-1 photon and the spin wave using purification, as done in Ref. 9; b) balancing the storage efficiency for different OAM modes by, e.g., smoothing the transverse distribution of the coupling laser beam in the cold atomic ensemble and the atomic density; c) reducing the background noise by using more strict filtering to achieve a better SNR; d) improving the working state of the whole system over long experimental periods.

There are many limiting factors for atomic storage time, including the residual magnetic field and atomic motion. In general, memory time can be improved by compensating for the magnetic field or by using magnetic field-insensitive states. By reducing atomic motion with an optical lattice, a millisecond, even a hundred millisecond storage time can be achieved. Moreover, the dynamic decoupling method can also be used to improve the storage time. In the present experiment, the storage time was also limited by the experimental time sequence, which was performed within hundreds of nanoseconds. The storage time can be improved further by optimizing the time sequence.

We emphasize that achieving high-dimensional entanglement between different quantum memories is non-trivial and is not a straightforward extension of establishing a two-dimensional entanglement between two atomic ensembles. There are many challenges both in its creation and verification. For distant memories entangled in a two-dimensional space [24], the entanglement can be well characterized by reconstructing the density matrix and checking the Bell-type inequality. In contrast, characterizing a high-dimensional entanglement is more complex and difficult. As noted earlier, reconstructing the density matrices using the method for two-dimensional entanglement is impractical for a high-dimensional entanglement (>3) because the amount of data that must be measured significantly increases. Therefore, we sought a different way to characterize it, using a witness instead. Moreover, it is not easy to balance the distinct efficiencies when storing different OAM modes. This is not a problem in the two-dimensional case because any two OAM modes with the same value but opposite sign serve as a two-dimensional space. Another significant difference is that a series of 4-F imaging systems have been designed subtly and constructed meticulously to detect the high-dimensional entanglement, instead of the direct projection onto SLM for a low azimuthal index of ±1 used in Ref. [24];

otherwise, we could not obtain the correct results for reconstructing the density matrix and calculating the witness. We believe these are the main reasons why there have been no reports of experimental progress until now. Indeed, in one important aspect, i.e., the storage of a high-dimensional OAM state on demand in any physical system and the establishment of a high-dimensional OAM entanglement between different quantum memories, our work represents primary progress and a significant step forward in this field.

Light-carried OAM cannot be transmitted in a commercial optical fiber; therefore, OAM-based quantum networks may be more suitable for work in a free space system. Recently, Zeilinger's group realized the distribution of OAM entanglement between two sites separated by 3 km in Vienna in 2015 [47]. We also note that light with OAM can transmit along some special fibers over kilometers [48]. Currently, many groups and people are working in this field; therefore, a quantum network based on OAM may be realized in the future.

Moreover, we note that Ref. 49 reported a quantum storage of a 3-dimensional entanglement in solid crystal. Compared to that work, in addition to the media for memory being different, our work reports a higher-dimensional entanglement storage. Most importantly, the memory we achieved can work on-demand; this is the key point for realizing a long-distance quantum communication based on a quantum repeater. The technique of a two-level atomic frequency comb used in that work cannot work on-demand; the storage time is predetermined. There is another difference between these two works: we achieved high-dimensional entanglement between two physical systems, not the OAM entanglement between a delayed photon and the atomic excitation achieved in solid storage.

## CONCLUSIONS

In summary, quantum memories entangled in high-dimensional space between two 1-m-distant atomic ensembles were experimentally established for the first time. The density matrices for the three-dimensional entanglement were reconstructed, giving 83.9%±2.9% entanglement fidelity. For a higher-dimensional case, we proved that at least a four-dimensional entanglement existed between two memories using an entanglement witness. After verifying the dimension of the entangled memories, the experimental data showed that there was a seven-dimensional entanglement within the two atomic-ensemble memories. The experiment to establish high-dimensional entangled memories is an important step towards high-dimensional quantum communications.

## ACKNOWLEDGMENTS

We thank Miles J. Padgett and Alison M. Yao for kindly helping us solve the different OAM phase superposition states. We also thank Yong-Jian Han for helpful discussions and Guo-Yong Xiang for loaning two SLMs. This work was supported by the National Fundamental Research Program of China (Grant No. 2011CBA00200) and the National Natural Science Foundation of China (Grant Nos. 11174271, 61275115, 61435011, and 61525504).

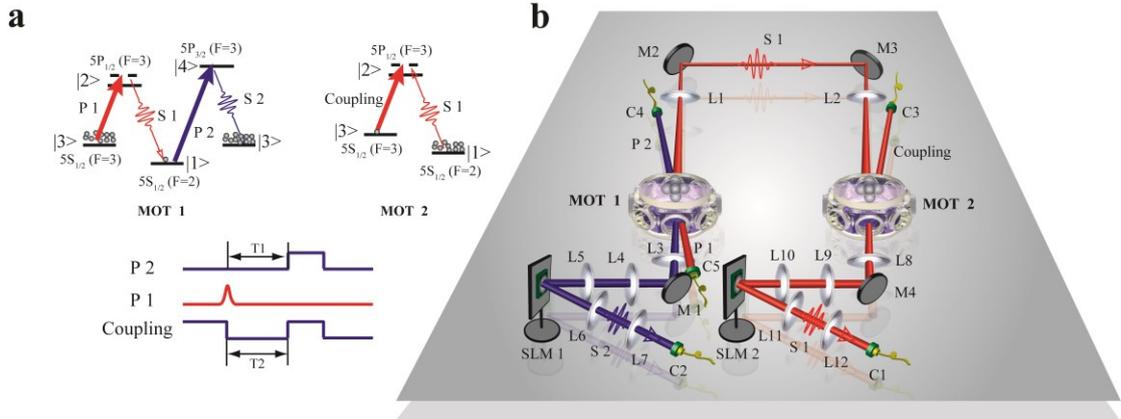

**Figure 1** (a) Energy level diagrams and the time sequence for creating and storing entanglement. (b) Experimental setup. Lenses L1 and L2 are used to focus signal 1 onto the center of MOT 2. L3, L4, and L5 are used to focus the phase structure of signal 2 onto the center of MOT 1 onto the surface of SLM 2. L6 and L7 are used to couple the OAM mode of signal 2 to C2. There is an asymmetric optical path for coupling signal 1 with C1 in the right half of the figure. P 1/2: Pump 1/2; S 1/2: Signal 1/2; C: fiber coupler; M: mirror; L: lens.

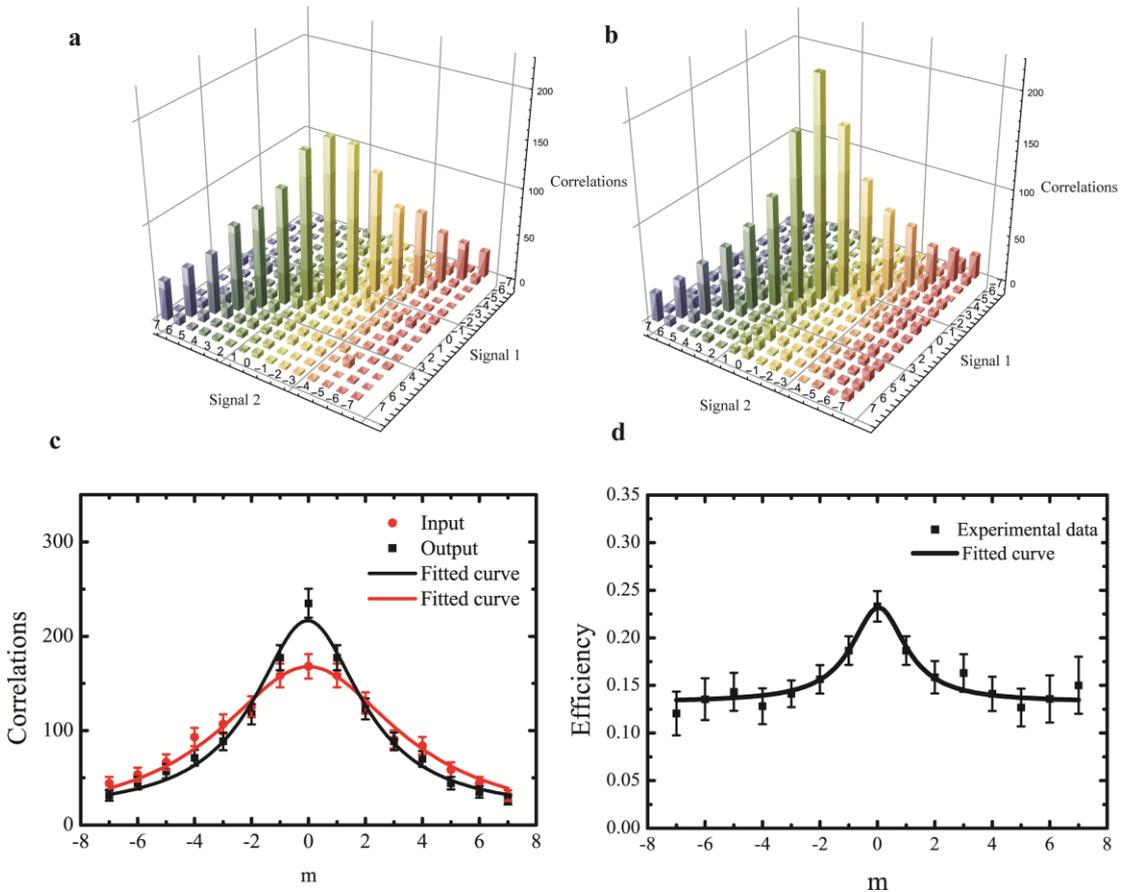

**Figure 2** Correlation between signal 1 and signal 2, with m=−7→7 before and after storage. (a) Coincidence rate before storage, measured over an interval of 100 s. (b) Coincidence rate after storage, measured over an interval of 600 s. (c) Distributions of the correlated OAM modes generated from SRS, where red dots and black squares represent datasets of input and output

OAM correlations, respectively. Both correlations are fitted using the fitting function $y = y_0 + \frac{2A}{\pi} \frac{w}{4(x-x_c)^2 + w^2}$, with ($y_0$=0, $x_c$=0, $w$=7.7, A=2030) and ($y_0$=12.7, $x_c$=0, $w$=4.57, A=1463), respectively. (d) The efficiency of storing different OAM modes. The black curve is calculated using the same fitting function, with fitted values ($y_0$=0.132, $x_c$=0, $w$=2.274, A=0.354); $w$ specifies the half-width at half-maximum of $y$. We identify $w$ with the effective quantum spiral bandwidth. For panels (c) and (d), the correlation refers to the coincidences of signal 1 and 2 photons, while the efficiency corresponds to the ratio of the coincidences before and after storage. Error bars represent ± s.d.

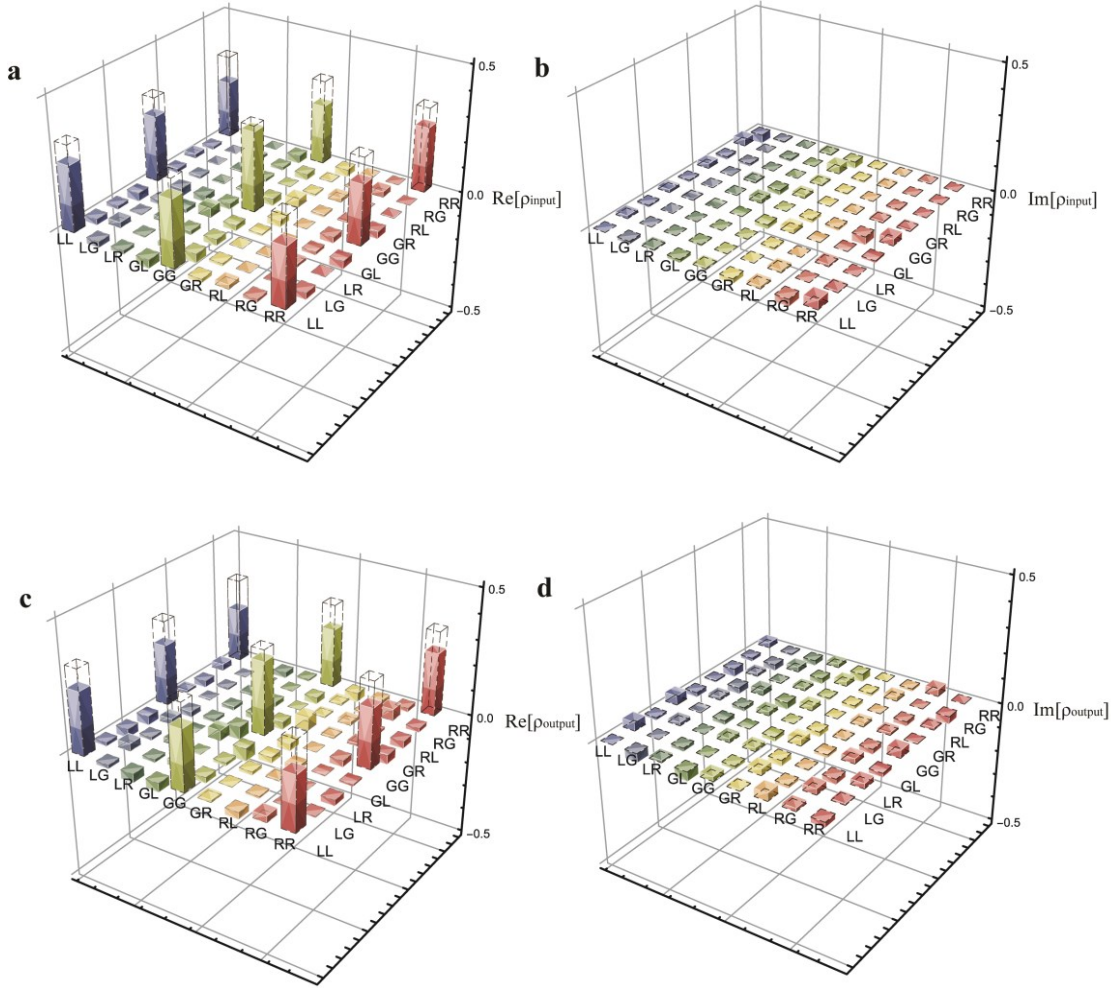

**Figure 3** Constructed density matrix of three-dimensional entanglement. (a) and (b) Real and imaginary parts before storage; (c) and (d) those after storage. The dotted bars added in each density matrix correspond to the expected value of the ideal density matrix.

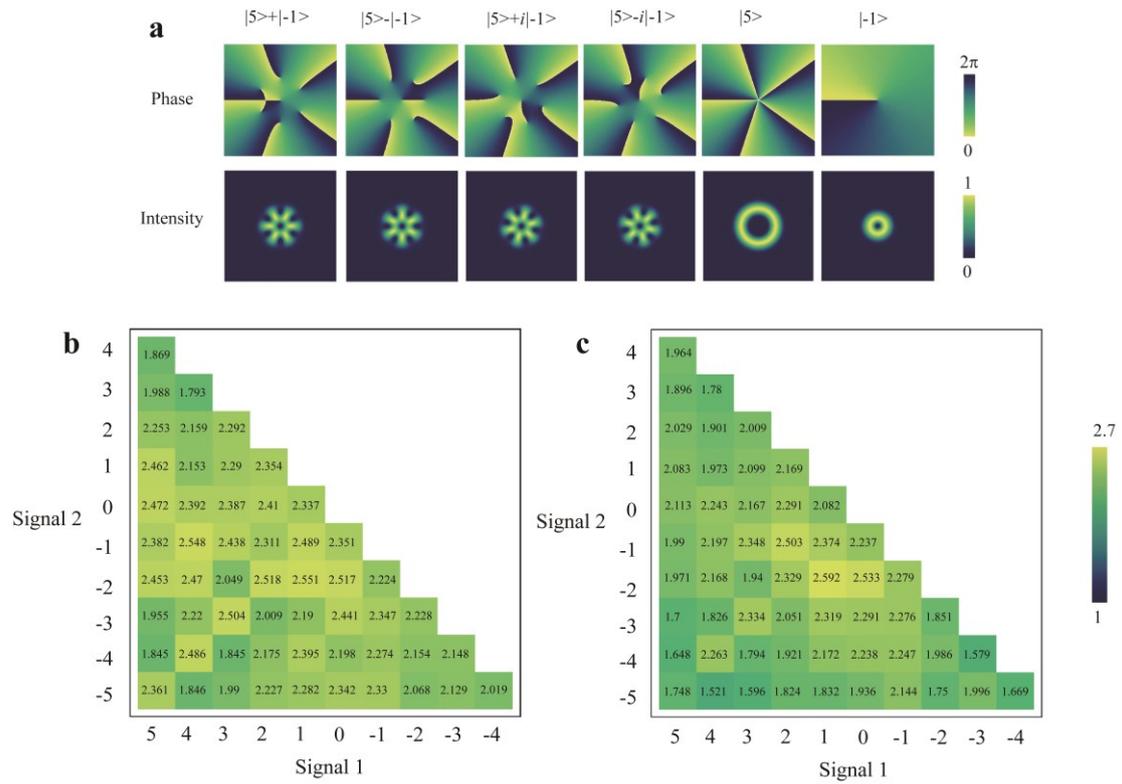

**Figure 4** (a) The diagonal/anti-diagonal, left/right, and horizontal/vertical bases in the phase and intensity spaces, with OAM modes $m$=5 and $m$=−1. The superposition is calculated by adding equal amounts of the two modes and the phase is calculated from the argument of the resultant complex [44], with the function Arg(LG$_5$+$e^{i\vartheta}$LG$_{-1}$), where LG$_5$ and LG$_{-1}$ are the amplitudes of OAM states, with azimuthal indexes of 5 and −1, respectively, and $\vartheta$ represents the relative phase. (b) and (c) are the visibilities before and after storage. A sum of the visibilities in three arbitrary OAM modes larger than six indicates the existence of two-dimensional entanglement.